\documentstyle[11pt,twoside,sympo2011,epsfig]{article}
\begin{document}
\title{Some CoRoT highlights - A grip on stellar physics and beyond}
\author{E. Michel$^1$, A. Baglin$^1$ and the CoRoT Team}
\affil{$^1$ LESIA, Observatoire de Paris, CNRS UMR 8109, Univ. Pierre et Marie Curie, Univ. Paris Diderot, pl. J. Janssen, 92195 Meudon, France
 [Eric.Michel@obspm.fr]} 
\begin{abstract}
About 2 years ago, back in 2009, the first CoRoT Symposium was the occasion to present and discuss
unprecedented data revealing the behaviour of stars at the micromagnitude level.
Since then, the observations have been going on, the target sample has enriched
and the work of analysis of these data keeps producing first rank results.

These analyses are providing the material to address open questions
of stellar structure and evolution and to test the so many physical 
processes at work in stars. 
Based on this material, an increasing number of interpretation studies
is being published, addressing various key aspects: the extension of
mixed cores, the structure of near surface convective zones, 
magnetic activity, mass loss,
... Definitive conclusions will require cross-comparison of 
results on a larger ground (still being built), but it is already possible at the time of this Second 
CoRoT Symposium, to show how the various existing results take place in a general framework
and contribute to complete our initial scientific objectives.
A few results already reveal the potential interest in considering stars and planets globally,
as it is stressed in several talks at this symposium.  
It is also appealing to consider the fast progress in the domain of Red Giants and
see how they illustrate the promising potential of space photometry beyond the
field of stellar physics, in connex fields like Galactic dynamics and evolution.

\end{abstract}
\section{High quality light curves for a wide selection of stars}
At the time of this Symposium, CoRoT has observed about 125 selected 'bright' stars (5.4$\le$m$_V\le$9.5)
with a 1s sampling time and a noise level dominated by the photon counting noise 
(approximately $3~10^{-4} < \sigma_{1s} < 2~10^{-3}$)  
and about 130000 fainter ones (11$\le$m$_V\le$16) with a 512s sampling time and a noise level about
2 times the photon noise ($3~10^{-4} < \sigma_{512s} < 3~10^{-3}$).

\begin{figure}[h]
\begin{center}
\epsfig{width=12cm,file=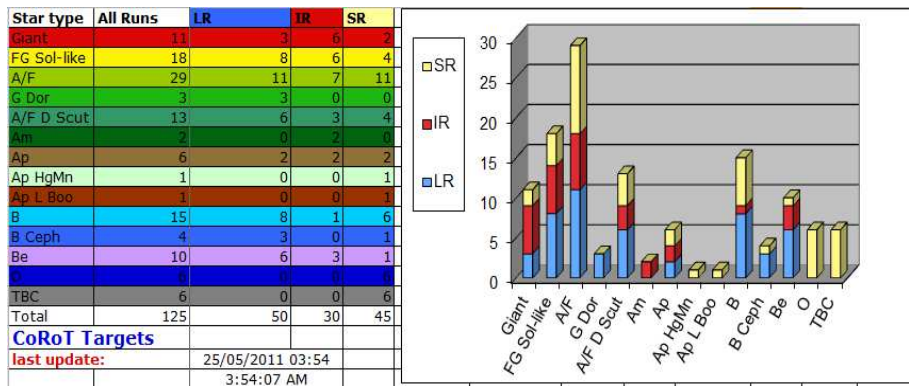}  
\caption{Distribution of the 'bright' stars selected and observed in the seismo field, in terms of star type and length of the runs}
\end{center}
\end{figure}

These observations are composed of Long Runs (140-170 days) for 40\%, Intermediate Runs (60-80 days)
for 24\% and Short Runs (20-30 days) for 36\% (see Fig.1). They all show a duty cycle larger than 90\%.

The stars observed cover a large variety of masses and evolution stages (see Fig.2) as planed in the original
scientific programme (see Michel et al. 2006).

\begin{figure}[h]
\begin{center}
\epsfig{width=12cm,file=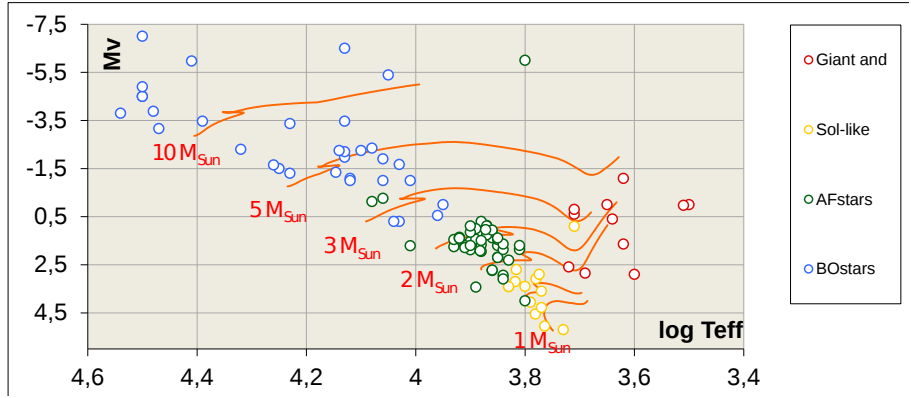}  
\caption{Distribution of the 'bright' stars selected and observed in the seismo field, in the HR diagram}
\end{center}
\end{figure}
\section{Precise characterization of oscillation modes and beyond}
A practical objective for CoRoT was to characterize stellar oscillations by measuring with precision mode parameters
(frequencies, amplitudes, lifetimes, ...). In this respect, the solar-type pulsators were the most challenging
and one of the first success of CoRoT was to perform measurements of mode parameters
for solar-type oscillations in stars other than the Sun. The first observations of early F stars
like HD~49933 and HD~181420 (see Appourchaux et al. 2008, Barban et al. 2009) revealed amplitudes only slightly lower
than expected, but lifetimes shorter by a factor 3. 
Since then, observations have been extended with succes to more evolved stars like HD49385 (see Deheuvels et al. 2010, and these proceedings),
but also to stars closer to the Sun in 
terms of mass and evolution stage
like HD~52265 (see Ballot et al. 2011, and these proceedings) or HD~43587 (Boumier et al. these proceedings) and recently
to HD~42618, a G type star with a large separation $\Delta \nu = 141 \mu$Hz, presently the closest to the Sun
for which individual modes are measured (see Fig.3). 

For self sustained ('opacity driven') pulsators, the CoRoT data allowed to characterize modes hundred 
times smaller in amplitude than
what was previously done, on time scales unaccessible so far (see e.g. Poretti et al., Briquet et al., Neiner et al., these proceedings).

Finally, for many stars accross the HR diagram, the CoRoT light curves revealed signature of phenomena other than oscillations,
like spots or granulation in F stars, but also probably in hot O stars (Blomme et al. 2011) and Red Giant stars (Carrier et al. 2009) 
and possibly in A stars (Kallinger \& Matthews 2010).

\begin{figure}[h]
\begin{center}
\begin{tabular}{cc}
\epsfig{width=5cm,file=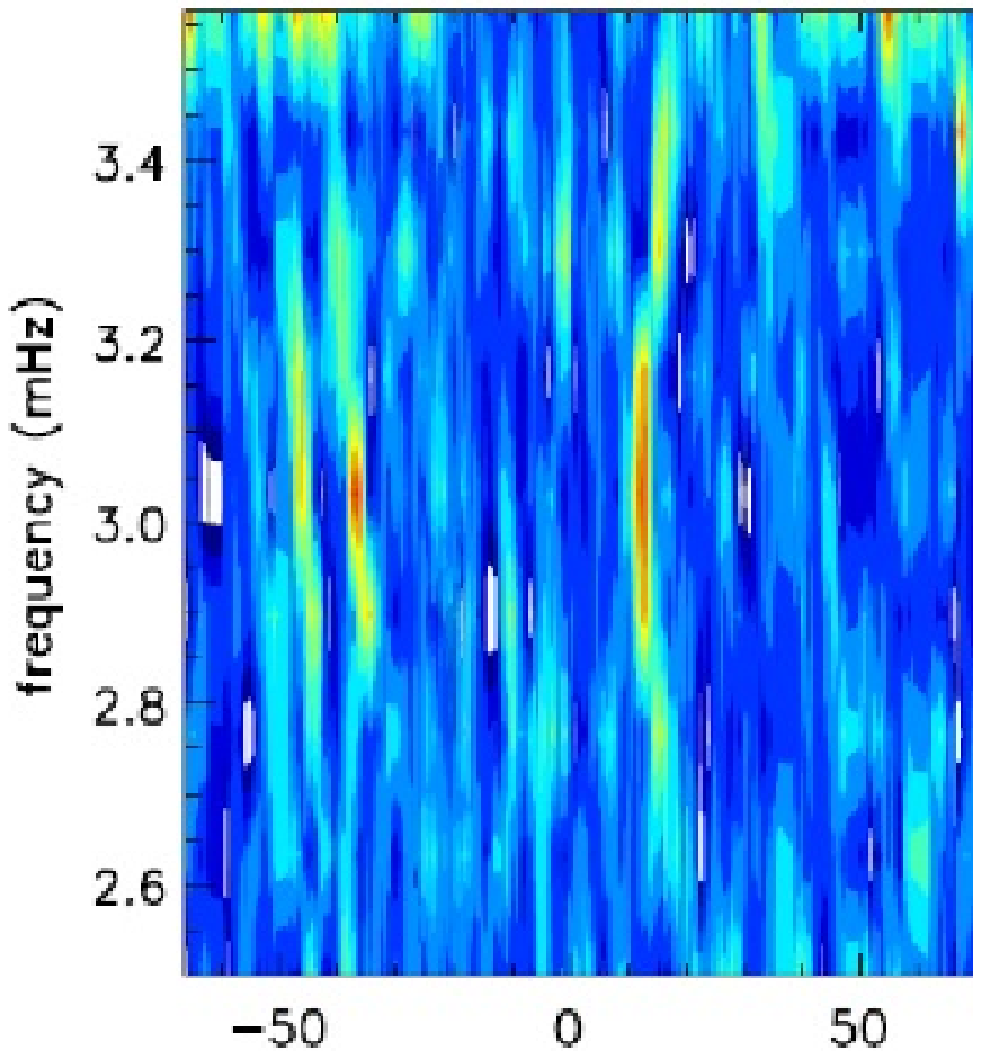} &
\epsfig{width=9cm,file=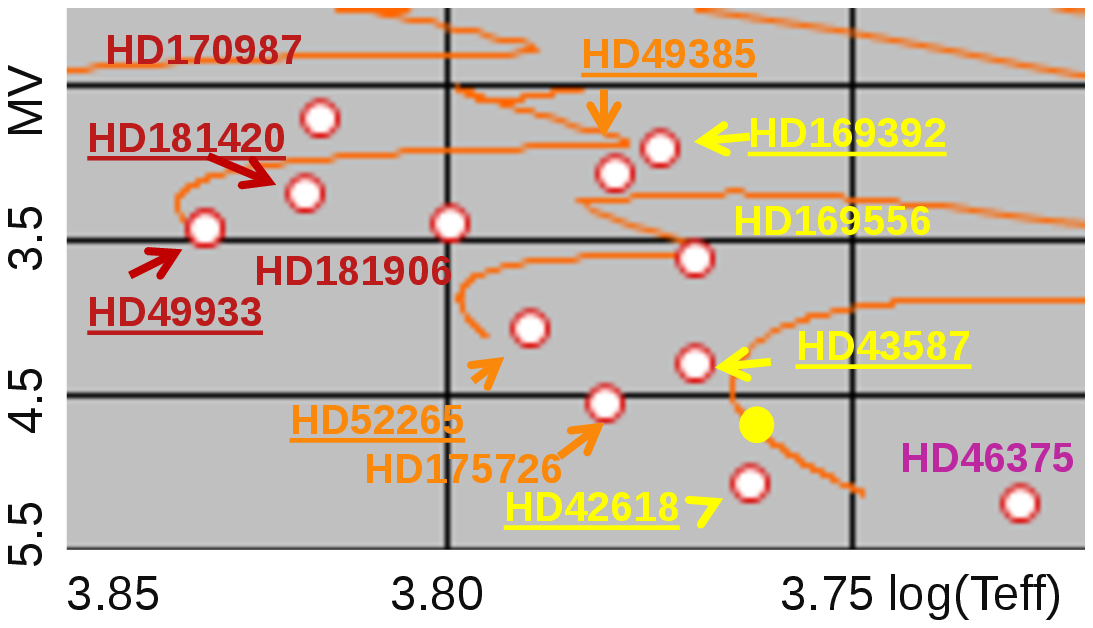} \\ 
\end{tabular}
\caption{left: Echelle diagram for HD~42618, a close neighbour to the Sun in terms of global parameters with
$\Delta \nu = 141 \mu$Hz and $\nu_{max}~3150 \mu$Hz. right: HR diagram showing solar type pulsators on or near the Main Sequence observed with CoRoT}
\end{center}
\end{figure}

\section{beyond the observational succes}
CoRoT has discovered signature of very diverse phenomena, some of them expected for 
long and some completely new.
Among the observational successes, one can remind the first detection and precise characterization
of solar-type oscillations in stars other than the Sun (Appourchaux et al. 2008, Michel et al. 2008)
which was expected for decades, the discovery of the underlying structure of Red Giants oscillations
(De Ridder et al. 2009) which openened the way to a brand new and vivid field for seismology,
the first detection of solar-like oscillations in massive stars (Belkacem et al. 2009, Degroote et al. 2010)
which are still subject of great debates and great expectations as well as the discovery of hundreds of
low amplitude peaks in delta Scuti stars (Poretti et al. 2009), the first precise follow up of the oscillations
of a Be star during an outburst (Huat et al. 2009), a unique view on mass-loss and pulsation relations, the first detection
of deviation in g-modes period spacing in a massive star (Degroote et al. 2010), a direct signature of the edge of the convective core,...

Beyond this technical and observational success, theoretical interpretations of CoRoT data are growing rapidly in number and in diversity. 
These studies have already addressed most of our scientific objectives as illustrated hereafter.
\section{Extension of mixed stellar cores}
Characterizing the extension of mixed cores, beyond the formally convective region
is a key objective in stellar physics because of the
role of tank this region plays for the nuclear reaction material and its
crucial impact on the evolution pace and stellar age. Several hydrodynamic processes 
are possibly involved and no quantitative
theoretical description exists. We thus need observational constraints for various stars, since this
phenomenon is expected to depend (at least) on mass and evolution stage, as well as on rotation.
Several studies address this point already (see Fig.4). 

\noindent
{\bf Stars with mass slightly higher than the Sun on or near the Main Sequence:}
The analysis of these data and their comparison to theoretical models suggest, in the case of the early F star HD~49933,
the existence of an extension distance $d_{ov}$ of the order of $0.1-0.2~$Hp 
(Hp the local pressure scale height) according to Benomar et al. (2010).  

\begin{figure}[h]
\begin{center}
\epsfig{width=14cm,file=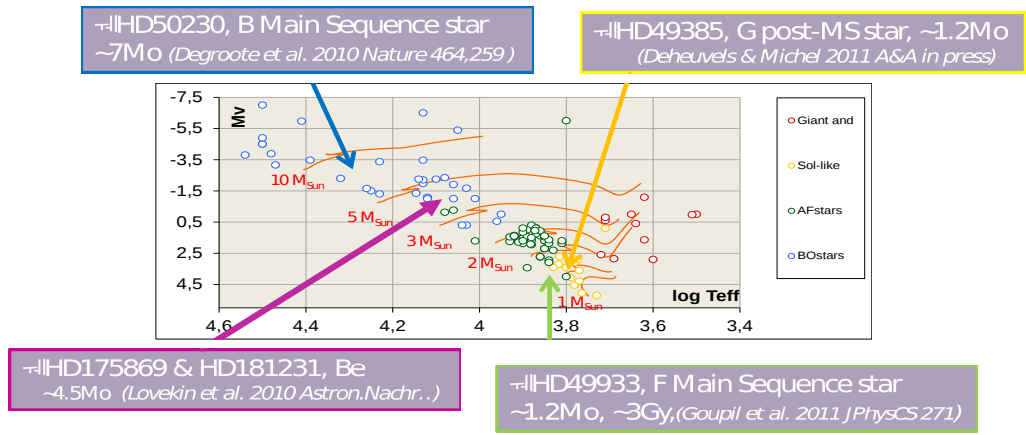} 
\caption{ 
Diagram HR and position of various objects mentioned in the text which study addresses the question of the mixed core extension.
}
\end{center}
\end{figure}

In the case of the cooler star HD~52265 on the other hand, first studies (see Escobar et al., these proceedings) suggest no need
for an extra mixing beyond the formal Schwarzschild limit, 
while for HD~49385, a $\sim$1.2M$_{\odot}$ G star just off the Main Sequence, 
seismic interpretation by Deheuvels \& Michel (2011) tends to exclude mild values but let open possibility for a large
or a very small extension $0.18~$Hp$ \le d_{ov} \le 0.2~$Hp or $d_{ov} \le 0.05~$Hp).

\noindent
{\bf More massive stars on the Main Sequence:}
In the $\sim$7M$_{\odot}$ B star HD~50230, the measurements of g modes revealed distribution in period which are 
interpreted as the signature of the sound speed variation associated with the gradient of chemical composition at the edge of the core
(Degroote et al. 2010). This study suggests a rather important extension of the mixed core ($d_{ov} \ge 0.2~$Hp).

The modeling of two $\sim$4.5M$_{\odot}$ late Be stars HD~181231 and HD~175869 (Neiner et al. A\&A in press) suggests the need for an even 
larger extension in terms of local pressure scale height ($d_{ov} \ge 0.3~$Hp)  
to match the observed frequencies with the prograde modes theoretically expected in these very fast rotators.
The same authors seize this opportunity to propose a quantitative discussion of the relative contribution
of the different processes possibly responsible for such an extension.
This work is illustrative of possible future applications gathering the differents indications we have and will 
obtain from various stars.

\section {The structure of the outer convective zones}

The outer convective zone present in stars with more or less extension just bellow the surface is another main source of uncertainty 
in our modeling and understanding of stars. 
It constitutes a poorly determined transition zone between what we see from the star at the surface and the internal part
that we 'interpolate' with theoretical models, structurally and chemically speaking. 

Two specific features of the convection zone associated with sound speed rapid variation, the He ionisation zone and 
the  bottom of convective zone,  manifest themselves through 
'oscillations' which can be seen in the large separation and second difference, as a function of frequency. 
These oscillations have been detected for the first time in the solar like star HD~49933 
(Mazumdar \& Michel 2010, and Mazumdar et al. these proceedings) and also in a bright giant HD 181907 (Miglio et al. 2010). 
Their periods give an estimate of the acoustic depth of these two discontinuities (Fig.5) , 
though the depth of the convective zone is more difficult to assess.

\begin{figure}[h]
\begin{center}
\begin{tabular}{cc}
\epsfig{width=7cm,file=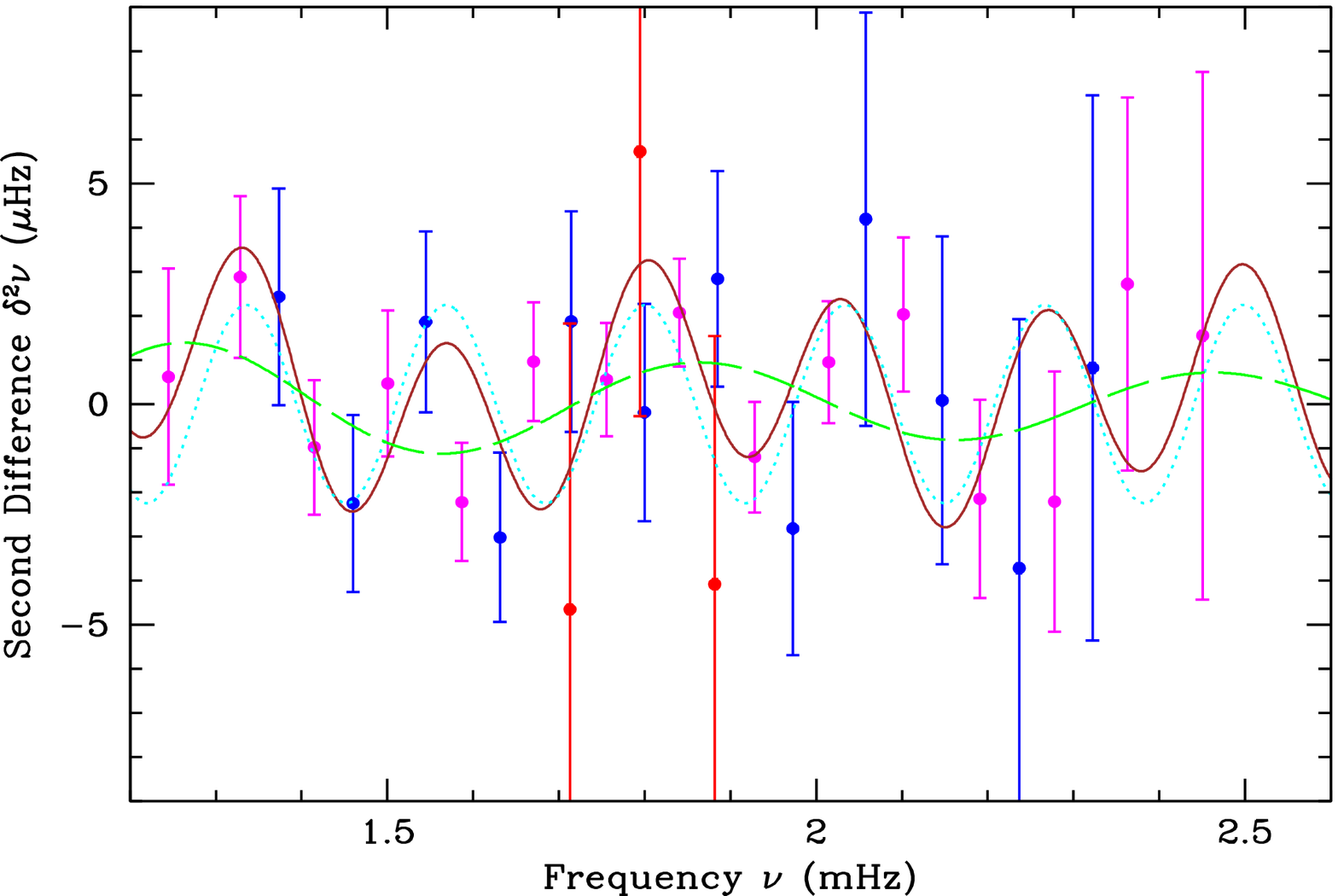} &
\epsfig{width=8cm,file=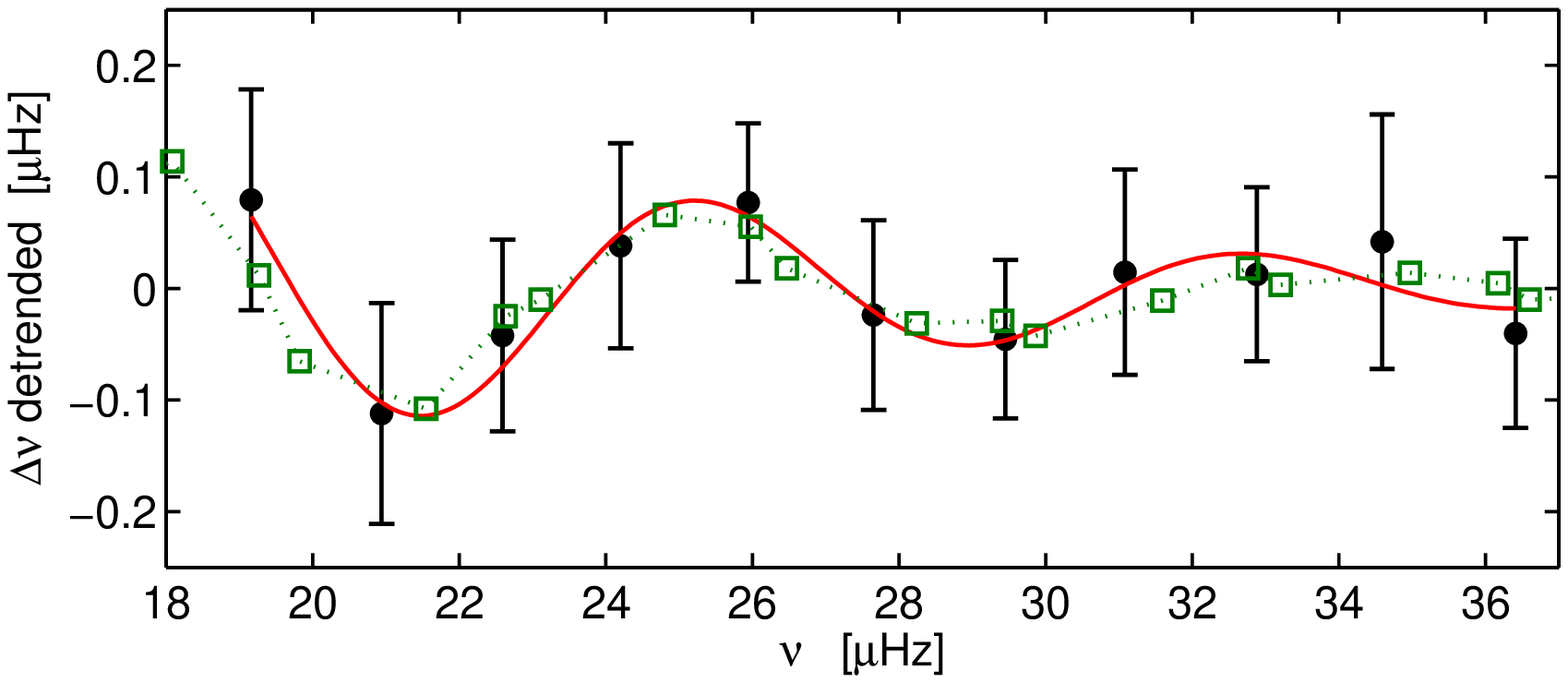} \\
\end{tabular}
\caption{ 
Detection of the oscillation of the second difference and of the large separation as a function of frequency, in HD 49933 a main sequence star 
{(\it left, from Mazumdar and Michel 2010)}, and in the giant star HR 7349 {\it (right, from Miglio et al. 2010)}. 
}
\end{center}
\end{figure}

\subsection {Super adiabatic outer layers}

Several signature exist in the data which can tell us about the convection in the superadiabatic layer where classical
descriptions by the mixing length fail to reproduce the energetics of the convective transport precisely enough.  

\noindent
{\bf Amplitudes and line widths:}
The seismic  data obtained with CoRoT  enable us for the first time  to measure individual amplitudes 
and line widths of solar-like oscillations for stars other than the Sun.
From those measurements it is possible, as was done for the Sun, to
constrain models of the excitation of acoustic modes by turbulent convection.
Using the seismic determinations of the mode line widths (benomar et al. 2009)
and the theoretical mode excitation
rates computed for this specific case (Samadi et al. 2010a), Samadi et al. (2010b) 
have derived the expected surface 
velocity amplitudes of the acoustic modes. 
Using a calibrated quasi-adiabatic approximation relating the mode amplitudes in intensity to those in velocity, 
they have finally derived the expected values of the mode amplitude in intensity.

First calculations assuming solar abundances  showed theoretical amplitude in excess by $\sim$~35\,\%.
When taking subsolar metal abundance of HD~49933 (Fig.6a), the theoretical amplitudes are within 1-$\sigma$  
error bars of the measured ones, except at high frequency.

\begin{figure}[h]
\begin{center}
\begin{tabular}{cc}
\epsfig{width=7cm,file=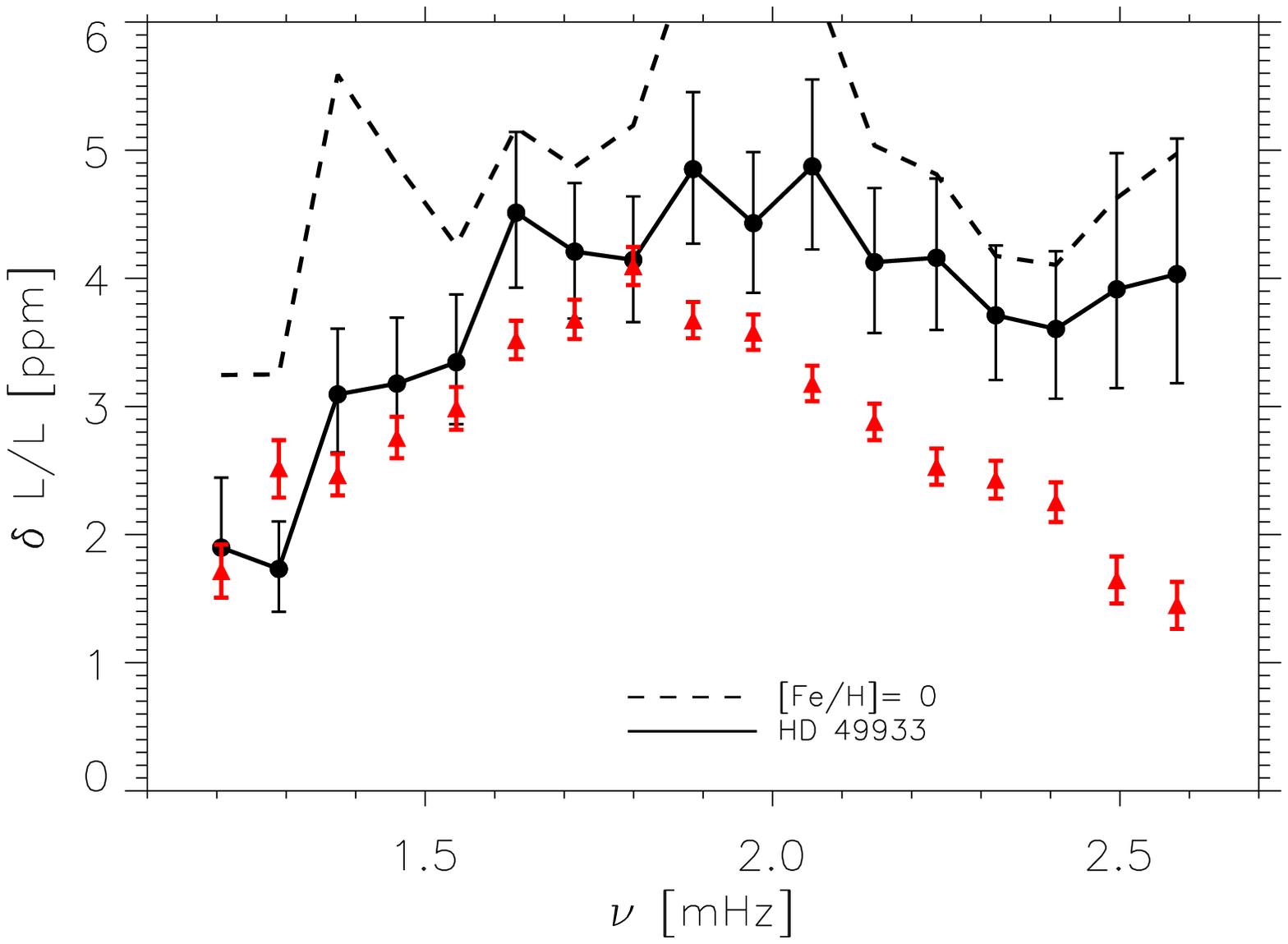} &
\epsfig{width=7cm,file=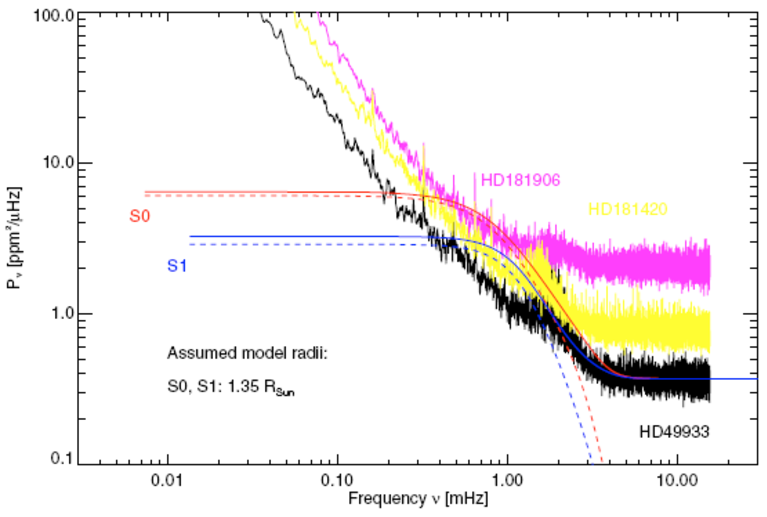} \\
\end{tabular}
\caption{ 
HD49933.
{\it left}:  Theoretical bolometric amplitude of the modes with solar (dashed line) and subsolar (solid line) 
metallicity compared with measured ones (red triangles), from Samadi et al.(2010b);  
{\it right}: Power density spectrum (PDS) measure for HD49933 (black) and compared with theoretical ones   
for solar metallicity (S0, in red) and strongly subsolar metallicity (see text, S1, in blue), from Ludwig et al.(2009).
}
\end{center}
\end{figure}

These results seems to validate the main assumptions of the model of stochastic 
excitation in the case of a star significantly hotter than the Sun. 
However, the discrepancies seen at high
frequency($\nu \ge$1.9~mHz) highlight some deficiencies of the modeling, whose origin remains to be understood.

\noindent
{\bf Granulation:}
HD 49933 shows also very clear evidence of a photometric granulation background (Michel et al. 2008).
Ludwig et al. (2009) have produced disk-integrated brightness fluctuations directly emerging from representative 3D hydrodynamical 
models of the atmosphere of HD49933: one with a solar metal 
abundance and the second with a metal to hydrogen abundance [Fe/H]=-1 (for HD~49933, [Fe/H]=-0.4). 
Both models have the same effective temperature and gravity, 
close to that of the target.
As shown in Fig.6b, the calculation results in a significant over-estimate in total power.

The level of the granulation background significantly decreases with surface metal abundance,
but remains higher than the observations by a factor 2.

At the present time, we are then left with a puzzling discrepancy between the predicted and observed granulation background in HD 49933, 
and this is confirmed by several other objects in this domain.

\section {Magnetic activity and rotation}

Rotation and magnetic field are two additional phenomena which influence on structure and evolution is poorly understood while it is
suspected to be important. 

\subsection {Stellar cycles detected from modes parameters}

Using the CoRoT observations of HD 49933 separated by one year, Garcia et al.(2010) have detected for the first time the signature of
a magnetic cycle in the variation in frequency and amplitude of the modes. This  behavior (Fig.7a) is well known in the Sun. 
The present observations suggest a period of the order of 120 days for this star which is rotating about 8 times faster than the Sun. 
These results have not been reproduced so far on other objects (see Mathur et al., these proceedings).

\begin{figure}[h]
\begin{center}
\begin{tabular}{cc}
\epsfig{width=7cm,file=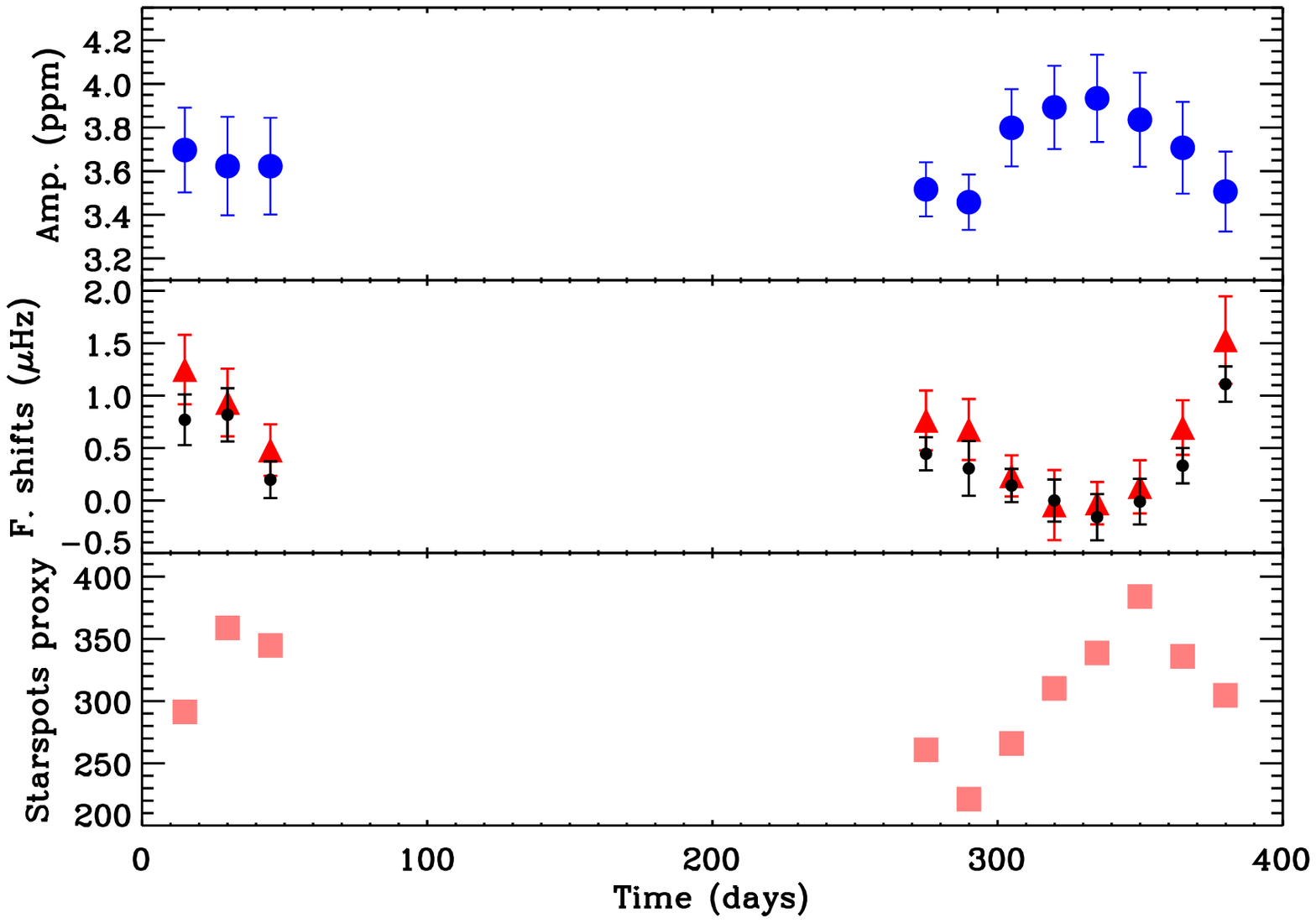} & 
\epsfig{width=7cm,file=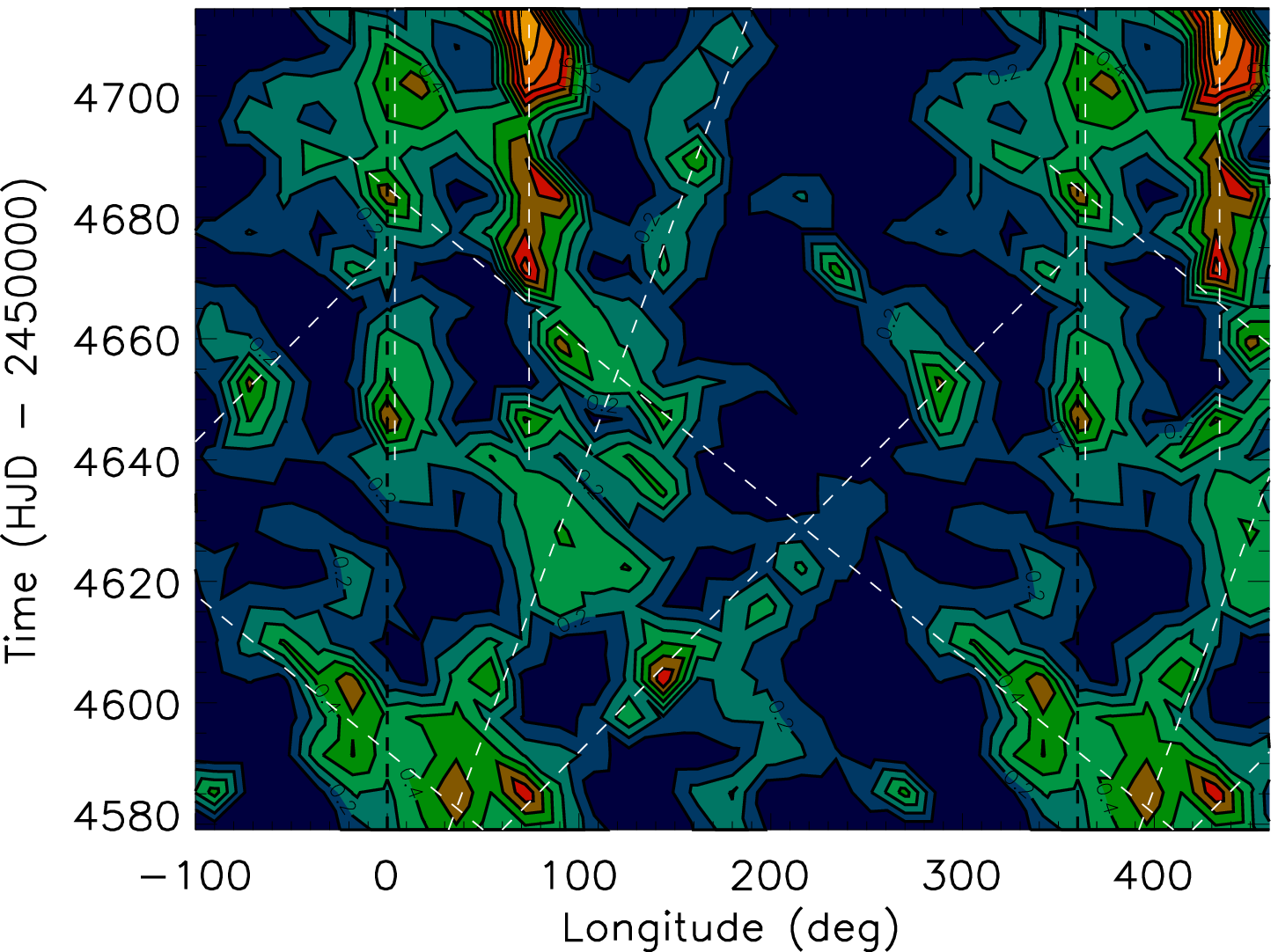} \\ 
\end{tabular}
\caption{ 
{\it Left}: 
The stellar cycle of HD 49933: Evolution with time of the mode amplitude (top), 
the frequency shifts (middle), measured with cross correlations (red triangles) and individual 
measurements (black circles), and a starspot proxy (bottom) built by computing the standard deviation of the light curve. 
{\it from Garcia et al. 2010} .
{\it Right}: 
Spot modeling and differential rotation of CoRoT 6 from Lanza et al.(2011).
Isocontours of the ratio f / fmax, where f is the spot covering
factor and fmax = 0.0059 its maximum value, versus time and longitude
The contour levels are separated by
0.1 fmax with yellow indicating the maximum covering factor and dark
blue the minimum. The dashed white lines trace the migration of the
active regions associated with each active longitude.
}
\end{center}
\end{figure}

\subsection {Spot modeling and surface differential rotation}
The first long duration light curves of CoRoT have triggered an intense activity in terms of surface modeling, 
identification of hotter and cooler regions at the surface and their variation with time (Fig.7b).
The first very successful attempt concerns the host star of the planet CoRoT-2b, an active G7 dwarf 
(Lanza et al. 2009, Silva-Valio et al. 2010). Different techniques have been used, giving coherent results (Savanov, 2010). 
Since then many other objects have been studied, starting to give some hints on the distribution of spots and the area covered, 
as well as indications of star-planet interaction.

Improvements of the modeling can be obtained using  complementary spectroscopic observations. 
This has been done for CoRoT 7 where a very intensive follow-up programme has been achieved to 
confirm the planetary nature of the transiting object. Lanza et al. (2010) for the first time 
compared simulated apparent radial velocity changes  induced by the distribution of active regions 
derived from the light curve modeling, to the spectroscopic observations. 
They show that magnetic activity cannot be responsible for a longer period oscillation around 3.7 days, 
and comforts the hypothesis of a second orbiting object in the system.

\subsection {Activity indices}

A first attempt to define activity indices from the analysis of light curves has been proposed by Hulot et al. (2011).
The low frequency component is generally sufficiently accurate to measure a reliable activity index, which captures only the low frequency energy excess.
When a rotational modulation is seen in the light curve, a Rossby number can be computed, using a theoretical estimate of the convective turnover time.
Figure 8 shows a clear trend of a decreasing  activity index with Rossby number; the large scatter is probably due to different physical situations as for instance differential rotation.

\begin{figure}[h]
\begin{center}
\epsfig{width=8cm,file=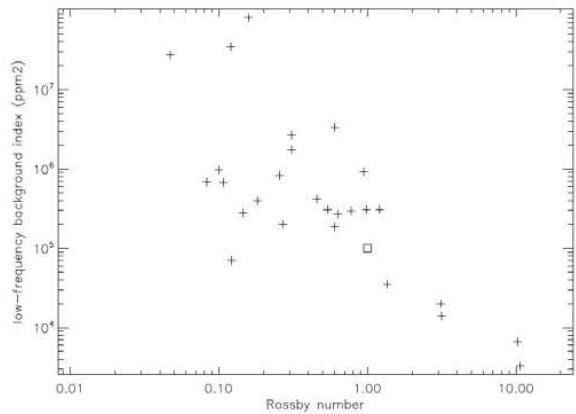} 
\caption{ 
The low-frequency background power index, or micro-variability
index, is plotted against the Rossby number. The active Sun, i.e. at its
maximum, corresponds to the square, while the quiet Sun would have a
log-index lower by 2. 
}
\end{center}
\end{figure}

\subsection {Star-planet connection}

As stressed in several talks at this conference, there is a practical interest in considering 
both planest and their central star globally. On one hand this can help to improve the characterization
of the system. This is illustrated already in the case of HD~52265 where the analysis of the oscillations modes
allow to provide an estimate of the inclinaison (Ballot et al 2011), but also in the HD~46375 (Gaulme et al. 2010)
where the simple measurement of the large separation allows to improve the estimate of the stellar mass and therefor
of the planet's mass. Beyond this practical aspect, stars and planets being born together, several interogations exist 
about the possible interactions they might have in their mutual history, as attested in several talks at this symposium.

\section {Stellar population studies}

As seen on  Figure 9, CoRoT and Kepler do not observe the same regions of the sky. The {\it  centre field } 
of CoRoT is not very far from Kepler one, though closer to the galactic plane. The {\it anticentre field} 
of CoRoT is  on the opposite direction, in a quite young surrounding.

This interesting situation is starting to give a strong push to population studies. 
A new branch of seismology {\it ensemble seismology} is born. It uses the seismic parameters 
to characterize stars and to understand their evolution stage. 
Let's mention that these parameters are very precise and almost independant of the distances of the object.

The very first CoRoT long run boosted the number of known pulsating Red
Giants from less than 10 to over 700 (Hekker et al. 2009). It
also established the existence of non-radial long lived mixed-modes and at
the same time the clear apearence of the p-modes large separation
signature in the observed spectra (de Ridder et al. 2009).
The fact that the  spectrum and the echelle diagram is
dominated by a p-modes structure characterizing mostly the envelope, very
much as  for the Sun,  has been established (Mosser et al. 2011). This already allowed impressive applications as illustrated by
Miglio et al (2009) who showed that the population of red
giants observed with CoRoT and characterized in the plane $\nu_{max}$ and $\Delta
\nu$ could be compared with theoretical expectations and suggest a regular star
formation rate rather than recent star burst events for this sample.

\begin{figure}[h]
\begin{center}
\begin{tabular}{cc}
\epsfig{width=8cm,file=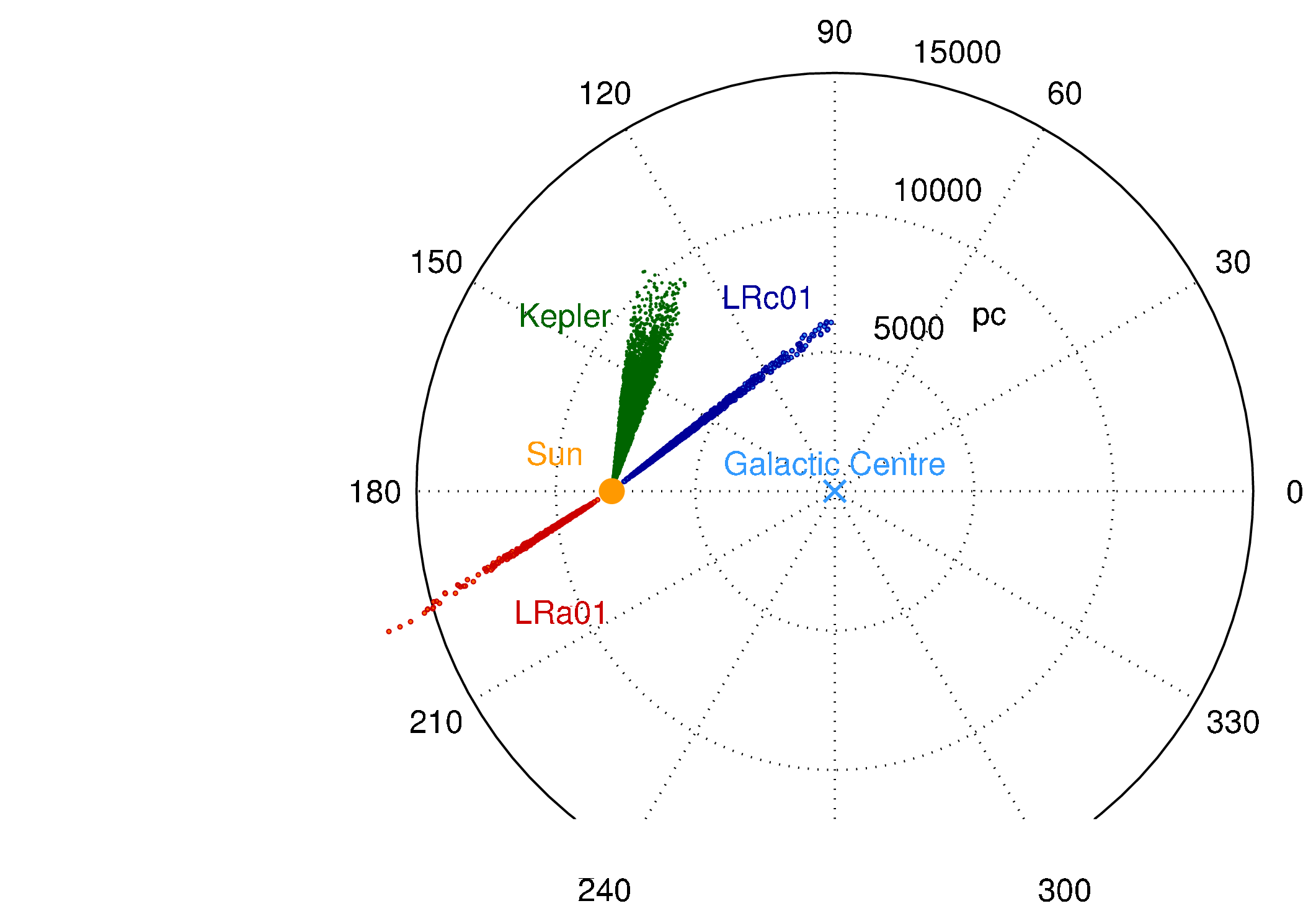} &
\epsfig{width=8cm,file=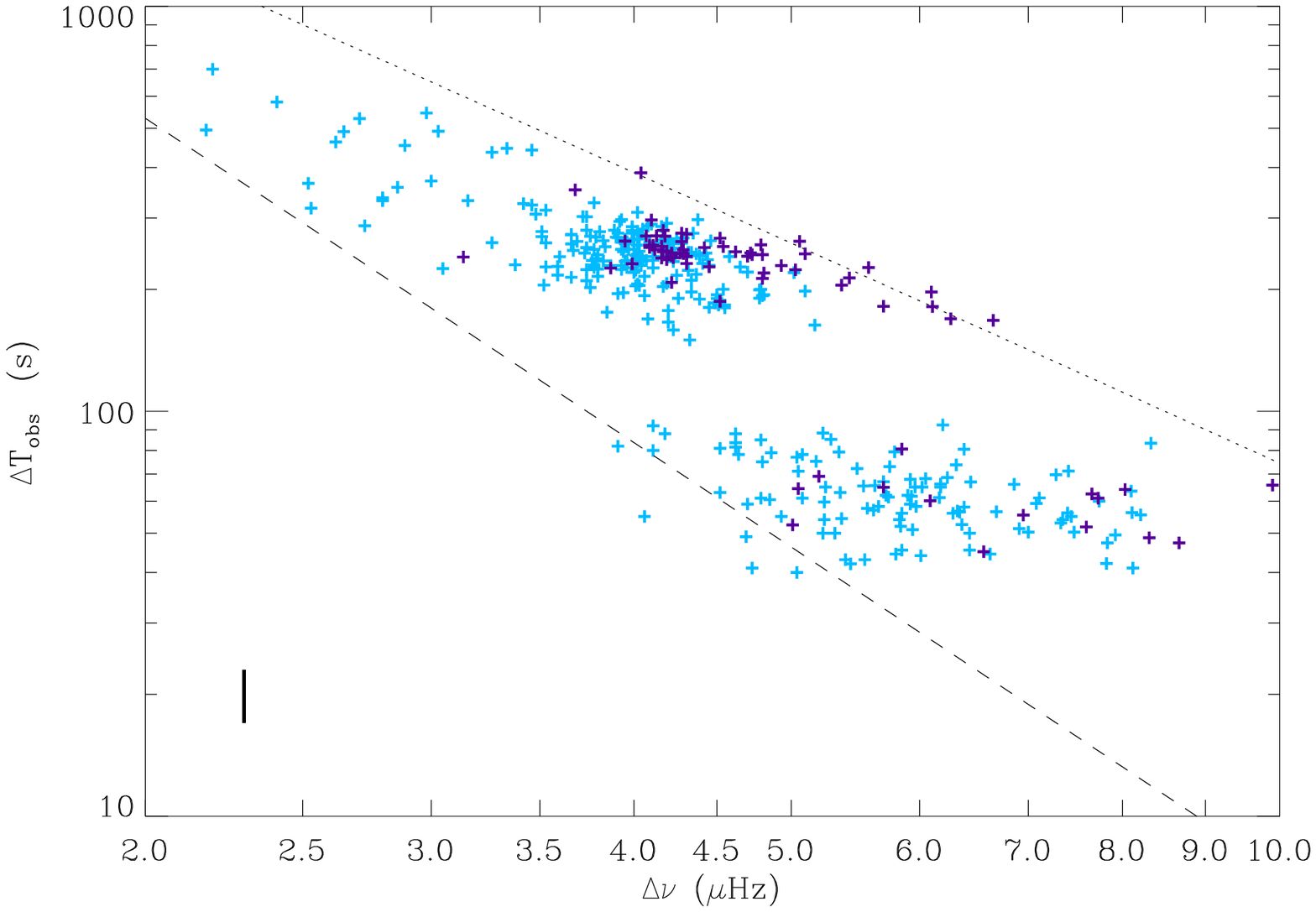} \\
\end{tabular}
\caption{ 
{\it Left}: 
Position in the Galactic plane of the giant stars observed with CoRoT in two different runs, 
LRa01 in the "anticentre" direction (red) and
LRc01 in the "centre" direction (blue) (sketch by A. Miglio), and with Kepler (green). 
{\it Right:} 
Diagram $(\Delta T/ \Delta \nu)$ for l=1 mixed modes in the two opposite regions observed by CoRoT:
centre (light blue), anticentre (blue). 
The upper cloud corresponds to stars in the helium burning core phase, 
while the lower one concerns stars in the hydrogen shell burning stage. 
{\it from Mosser et al. 2011}.
}
\end{center}
\end{figure}

Then, the pattern of mixed-modes characterizing the stellar core was also
revealed in Kepler and CoRoT data (Beck et al. 2011; Mosser et al. 2011).
 As shown by Bedding et al. (2011) and by Mosser et al. (2011), this pattern depends on the evolutionary stage. 
The differences are detected in the seismic data and then  allow to
discriminate red giants burning hydrogen in shell (Red Giant Branch) from
those burning helium in the core (Red Giant horizontal branch), which are undistinguishable from their surface fundamental parameters.

In addition the two directions of CoRoT observations allow to compare population of Red giants in different regions (Fig.9)

\subsection {Conclusion}

The CoRoT data have given a strong push to stellar seismology and the domain is gaining rapidly in maturity. As anticipated,
they allow to address open 
questions of stellar physics in a new light. 
It is also very appealing to see how these data may find applications in various connex domains beyond stellar physics. 
Among the first most obvious ones, the link between stars
and planets problematics  and the potential contribution to Galactic population study have been stressed in the programme of this symposium.

\end{document}